\documentclass[preprint,prl,superscriptaddress,prbib,11pt]{revtex4-1}
\usepackage{graphicx}
\usepackage{bm}
\usepackage{amsmath}
\usepackage{amssymb}
\usepackage{hyperref}
\usepackage{todonotes}
\usepackage{nicefrac}
\usepackage{bbm}
\usepackage[normalem]{ulem}
\usepackage{booktabs}
\usepackage{subfigure}
\usepackage{rotating}
\usepackage{color}
\usepackage{float}
\usepackage[T1]{fontenc}
\usepackage[utf8]{inputenc}
\begin{document}

\title{Spectroscopic fingerprints of a ferroaxial charge density wave}

\author{Jiangchang Zheng}
\thanks{These authors contributed equally.}
\affiliation{Department of Physics, The Hong Kong University of Science and Technology, Clear Water Bay, Kowloon, Hong Kong SAR}

\author{Zhongyi Zhang}
\thanks{These authors contributed equally.}
\affiliation{Department of Physics, The Hong Kong University of Science and Technology, Clear Water Bay, Kowloon, Hong Kong SAR}

\author{Fazhi Yang}
\thanks{These authors contributed equally.}
\affiliation{Department of Physics, The City University of Hong Kong, Kowloon, Hong Kong SAR}

\author{Josh Leeman}
\affiliation{Department of Chemistry, Princeton University, Princeton, NJ, USA}

\author{Luanjing Li}
\affiliation{Department of Physics, The Hong Kong University of Science and Technology, Clear Water Bay, Kowloon, Hong Kong SAR}

\author{Zihan Lin}
\affiliation{Department of Physics, The City University of Hong Kong, Kowloon, Hong Kong SAR}

\author{Zijian Fei}
\affiliation{Department of Physics, The City University of Hong Kong, Kowloon, Hong Kong SAR}

\author{Tianhao Guo}
\affiliation{Department of Physics, The City University of Hong Kong, Kowloon, Hong Kong SAR}

\author{Siyu Heng}
\affiliation{Department of Physics, The City University of Hong Kong, Kowloon, Hong Kong SAR}

\author{Xin Liang}
\affiliation{Department of Physics, The City University of Hong Kong, Kowloon, Hong Kong SAR}

\author{Leslie M. Schoop}
\affiliation{Department of Chemistry, Princeton University, Princeton, NJ, USA}

\author{Junzhang Ma}
\email{junzhama@cityu.edu.hk}
\affiliation{Department of Physics, The City University of Hong Kong, Kowloon, Hong Kong SAR}

\author{Hoi Chun Po}
\email{hcpo@ust.hk}
\affiliation{Department of Physics, The Hong Kong University of Science and Technology, Clear Water Bay, Kowloon, Hong Kong SAR}

\author{Berthold J\"ack}
\email{bjaeck@ust.hk}
\affiliation{Department of Physics, The Hong Kong University of Science and Technology, Clear Water Bay, Kowloon, Hong Kong SAR}

\date{\today}

\begin{abstract}
Unconventional charge density waves (CDWs) with complex order parameters can host exotic collective modes and non-trivial topologies. They have emerged as a new frontier in the study of quantum matter. Recent experiments on rare-earth tritellurides have reported evidence for a ferroaxial CDW through the detection of characteristic Raman modes. This phase, often regarded as a hidden order, has been recognized to arise from the coupling between charge and orbital degrees of freedom in these materials. Yet, spectroscopic insight into its underlying electronic structure and the explicit form of its order parameter symmetry has remained elusive. Here, we present results from linearly polarized angle-resolved photoemission spectroscopy (ARPES) and scanning tunneling microscopy (STM) measurements of the CDW phase in LaTe$_3$. Our ARPES measurements reveal a complex landscape of spectral gaps across the reconstructed Fermi surface, while our STM-based quasiparticle interference (QPI) mapping, enhanced through the selective deposition of atomic scattering centers, directly reveals an inter-orbital CDW with mixed $p_x$-$p_z$ orbital character. The detailed analysis of the QPI characteristics in terms of the order parameter symmetry within the orbital subspace of the Fermi surface suggests a mixed CDW phase with substantial ferroaxial component, which breaks all vertical mirror symmetries. More broadly, our work establishes a powerful spectroscopic pathway, based on scattering off individual atoms, for identifying and characterizing hidden, multi-component electronic orders in quantum materials using STM and ARPES measurements. %More broadly, the insights gained in this study thus pave the way for a deeper understanding and eventual optical control of ferroaxial order towards above room temperature data storage applications using tritellurides and other materials.
\end{abstract}

\maketitle

\subsection{Introduction}

{\bf [P1]} The search for unconventional electronic orders in quantum materials is a central theme in condensed matter physics, offering a gateway to new states of matter and emergent phenomena. Charge density waves (CDWs), which involve a periodic modulation of the electron density, have long served as a foundational paradigm for spontaneous symmetry breaking~\cite{gruner1988dynamics}. While the effects of conventional, single-component CDWs on electronic structure are well-established~\cite{zhu2015classification,gruner1988dynamics}, recent attention has shifted towards exotic CDWs characterized by complex, multi-component order parameters~\cite{li2021observation,aishwarya2023magnetic,singh2025ferroaxial, jiang2021unconventional, mielke2022time, guo2022switchable, denner2021analysis, feng2021chiral, zheng2025quasiparticle, gui2025probing}. Such states can break additional symmetries beyond those broken by the CDW wave-vector, potentially hosting novel collective excitations and non-trivial topologies, yet their direct experimental identification and characterization remain a significant challenge.

{\bf [P2]} The rare-earth tritellurides ($R$Te$_3$) have recently emerged as a key platform for exploring this new frontier~\cite{maklar2021nonequilibrium,yumigeta2021advances}. Initially studied as model systems for conventional CDW physics arising within the characteristic tellurium square lattice [see Fig.~\ref{fig:fig1}(a)], recent reports of an axial Higgs mode within the CDW phase of LaTe$_3$ and GdTe$_3$ using Raman spectroscopy~\cite{wang2022axial} suggest an unconventional CDW order, possessing a finite pseudo-angular momentum~\cite{wang2022axial,kogar2020light}. Subsequent studies on other members of the $R$Te$_3$ family~\cite{yumigeta2021advances,brouet2008angle}, such as ErTe$_3$~\cite{eiter2013alternative,moore2008fermi} and HoTe$_3$~\cite{pfuner2010temperature}, proposed the axial Higgs mode as the signature of a ferroaxial order~\cite{singh2025ferroaxial}, which is believed to arise from an intricate coupling between charge and orbital degrees of freedom in the CDW phase~\cite{hu2014coexistence, alekseev2024charge,que2025visualizing,singh2025ferroaxial,yumigeta2021advances}. This exotic state is an example of a 'hidden' order, because it couples linearly neither to magnetic nor electric fields~\cite{newnham2004properties}, making it a promising candidate for realizing non-volatile optoelectronic data storage devices~\cite{zeng2025photo}.

{\bf [P3]} Schematically, a ferroaxial CDW can be understood as a rotational, vortex-like arrangement of lattice distortions that collectively define a macroscopic axial vector. Its resulting order parameter breaks all vertical mirror symmetries (i.e., those parallel to the axial vector). However, it preserves both spatial inversion and time-reversal symmetry—the defining hallmark of a ferroaxial phase~\cite{newnham2004properties}. Theoretical models suggest that in the rare-earth tritellurides, this ferroaxial state naturally emerges from an `inter-orbital' CDW. This phase couples electronic states at crossing points of $p_x$- and $p_z$-orbital derived bands at the Fermi surface, generating an internal order parameter structure within the $p_x-p_z$ orbital subspace that inherently breaks all vertical mirror symmetries $m_x,\,m_y,\,m_{xy},$ and $m_{x\bar{y}}$ [see Fig.~\ref{fig:fig1}(b)]~\cite{hu2014coexistence, alekseev2024charge,singh2025ferroaxial}. To date, experimental evidence for a ferroaxial CDW in this material family is rare. It has been primarily inferred from collective mode signatures via Raman spectroscopy~\cite{wang2022axial,singh2025ferroaxial,chen2019raman}, structural symmetry analysis via electron diffraction~\cite{yumigeta2021advances,siddique2024realignment,ru2008effect} and second-harmonic generation~\cite{alekseev2024charge,singh2025ferroaxial}. However, a direct spectroscopic visualization of how ferroaxial order reconstructs the Fermi surface has remained elusive. This gap in our understanding is further compounded by recent strain-dependent measurements of ErTe$_3$ whose results challenge the presence of a ferroaxial phase~\cite{freitas2026revealing, singh2024emergent}. Indeed, the two-orbital subspace of the Fermi surface generally permits both conventional and ferroaxial CDW order parameter structures, depending on how $p_x$ and $p_z$ bands are coupled. This growing controversy underscores the critical and immediate need for direct, momentum- and real-space spectroscopic insights into the electronic structure and symmetry of the CDW states in these materials.

%{\bf [P3]} These findings have opened a novel chapter in the study of unconventional order in charge density waves, but they also highlight a critical gap in our understanding of ferroaxial CDWs. To date, the evidence for a ferroaxial CDW state has been primarily inferred from its collective mode behavior (via Raman spectroscopy)~\cite{wang2022axial,singh2025ferroaxial,chen2019raman}, lattice symmetry analysis (via electron diffraction)~\cite{yumigeta2021advances,siddique2024realignment,ru2008effect}, and the breaking of crystal symmetries (via second-harmonic generation)~\cite{alekseev2024charge,singh2025ferroaxial}. While theoretical analyses suggest that in rare-earth tritellurides ferroaxial order arises from an inter-orbital CDW phase composed of $p_x$ and $p_z$ bands at the Fermi surface~\cite{hu2014coexistence, alekseev2024charge,singh2025ferroaxial}, direct insights into how a ferroaxial charge order parameter is embedded in the reconstructed electronic Fermi surface has remained elusive to date. This leaves a fundamental question unanswered: what is the microscopic electronic texture of the ferroaxial CDW, and how does it manifest in the material's band structure and local density of states?

{\bf [P4]} Here, we directly probe the microscopic nature of the proposed ferroaxial CDW in LaTe$_3$. Angle-resolved photoemission spectroscopy (ARPES) measurements reveal distinct anisotropic gaps within the reconstructed Fermi surface. Quasiparticle interference (QPI) measurements with the scanning tunneling microscope (STM)—significantly enhanced through the selective 'in-situ' deposition of atomic-scale scattering centers—provide direct real-space evidence for the presence of an inter-orbital CDW phase. A symmetry analysis of our QPI maps within the orbital subspace of the Fermi surface reveals that this inter-orbital CDW is of mixed type with substantial ferroaxial component. Together, these results construct a microscopic picture of the electronic ground state of a ferroaxial CDW, bridging the gap between symmetry analysis, collective mode excitations and direct electronic-structure observation.

\subsection{Charge density wave state in LaTe$_3$}

\begin{figure}[H] 
    \centering
    \includegraphics[width=0.75\textwidth]{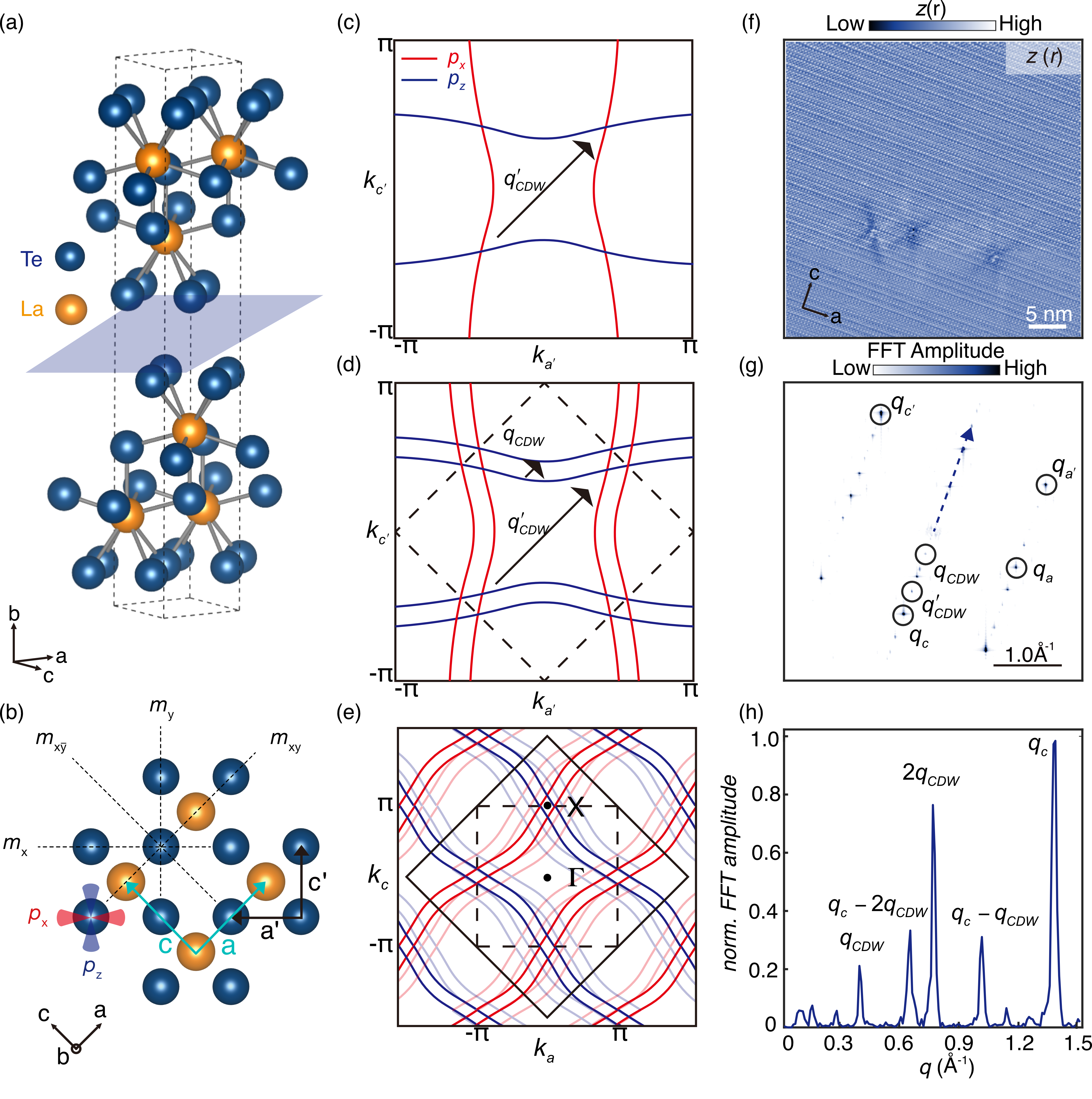}
    
     \caption{{\bf Charge density wave order in LaTe$_3$.} (a) Crystal structure of LaTe$_3$, with the blue plane indicating the cleavage plane. (b) Top view of the tellurium (Te) terminated surface and underlying lanthanum (La) layer; Te $p_x$ and $p_z$ orbitals are highlighted in red and blue color, respectively. The black dashed line marks the different vertical mirror symmetry planes (see side labels). Black arrows show the basis (\(a'\), \(c'\)) for one Te unit cell; turquoise arrows show the basis (a, c) for two Te unit cells. (c) Shown is the schematic band structure in the first Brillouin zone at the Fermi level for the single-Te primitive unit cell, showing the $p_x$- and $p_z$-dominated electronic states. (d) Shown is the schematic band structure at the Fermi level for the doubled two-Te unit cell. The dashed box outlines the corresponding reduced first Brillouin zone, upon enlarging the unit cell size. The unidirectional charge density wave (CDW) vectors $q_{\rm CDW}$ and $q'_{\rm CDW}$ are indicated. (e) Schematic presentation of the reconstructed band structure at the Fermi level with CDW order, showing the folded shadow bands with lighter color compared with original bands. (f) STM topography of the Te-terminated surface ($V_{\rm B}=300\,$mV, $I=500\,$pA, $T=4\,$K). (g) Two-dimensional fast Fourier transform (2D-FFT) of the topography shown in panel f. (h) Shown is a line cut along the dashed arrow in panel g, with peaks corresponding to $\mathbf{q}_\mathrm{CDW}$, $\mathbf{q}_c$, $\mathbf{q}_\mathrm{CDW}^\prime = \mathbf{q}_c - \mathbf{q}_\mathrm{CDW}$, and their linear combinations clearly labeled.}
    \label{fig:fig1}
\end{figure}

{\bf [P5]} LaTe$_3$, the lightest member of the rare-earth tritelluride (RTe$_3$) family, hosts a single CDW phase below a transition temperature T$_{\text{CDW}}$ $\approx$ 670 K~\cite{hu2014coexistence,yumigeta2021advances}. The material has an orthorhombic ($Cmcm$) layered crystal structure where weakly bonded, quasi square-lattice Te sheets with small in-plane anisotropy in the lattice constants ($a=0.999c$) are separated by strongly bonded La-Te blocks [Fig.~\ref{fig:fig1}(a)]. The van der Waals gap between these layers makes the Te-terminated surface the natural cleavage plane for experimental studies.

{\bf [P6]} The low-energy electronic structure of LaTe$_3$ is dominated by tellurium $p$-orbitals and a paradigmatic example of a square net compound. The in-plane Te $p_x$ and $p_z$ orbitals align along the $\mathbf{a}' = \mathbf{a}+\mathbf{c}$ and $\mathbf{c}' = \mathbf{a}-\mathbf{c}$ directions, respectively [Fig.~\ref{fig:fig1}(b)]. This orbital texture forms a quasi-2D Fermi surface composed of two nearly perpendicular sets of quasi-1D bands: $p_x$-derived bands dispersing along $\mathbf{k}_{c'}$ and $p_z$-derived bands dispersing along $\mathbf{k}_{a'}$ [Fig.~\ref{fig:fig1}(c)]. The presence of La atoms expands the primitive one-Te unit cell (black arrows, Fig.~\ref{fig:fig1}(b)) to a unit cell that accommodates two Te atoms (green arrows, Fig.~\ref{fig:fig1}(b)). This halves the Brillouin zone, folding the bands and creating a more complex Fermi surface comprised of eight distinct sheets in the reduced zone [Fig.~\ref{fig:fig1}(d)].

{\bf [P7]} As in other compounds of the RTe$_3$ family, the electronic structure of LaTe$_3$ is unstable towards the formation of a unidirectional CDW. Our STM measurements reveal a clear unidirectional modulation along the crystallographic $c$-axis, the hallmark of the incommensurate CDW state [Fig.~\ref{fig:fig1}(f)]. Both theory and experiment suggest that this CDW arises from a nesting instability, with a wave vector $\mathbf{q}_\mathrm{CDW}$ that connects the crossing points of the $p_x$ and $p_z$ bands~\cite{alekseev2024charge,sarkar2023charge} and could give rise to an inter-orbital CDW. In the folded two-Te Brillouin zone, this primary nesting vector is accompanied by a secondary vector $\mathbf{q}_\mathrm{CDW}^\prime = \mathbf{q}_c - \mathbf{q}_\mathrm{CDW}$, where $\mathbf{q}_c$ is the reciprocal lattice vector along the ordering direction~\cite{nakamura2024revealing}.

{\bf [P8]} The two-dimensional fast Fourier transform (2D-FFT) of our STM topography data confirms this picture, revealing a rich set of scattering vectors [Fig.~\ref{fig:fig1}(g)]. A line profile of the 2D-FFT [Fig.~\ref{fig:fig1}(h)] shows peaks that can be indexed to $\mathbf{q}_\mathrm{CDW}$, its counterpart $\mathbf{q}_\mathrm{CDW}^\prime$, and the Bragg peak $\mathbf{q}_c$. We measure $\mathbf{q}_\mathrm{CDW} = 0.276\,\mathbf{q}_c$, consistent with previous reports for incommensurate CDW in LaTe$_3$~\cite{sarkar2023charge}. The CDW formation reconstructs the pristine electronic structure by translating the bands by multiples of $\pm\mathbf{q}_\mathrm{CDW}$. The result is a complex Fermi surface featuring original bands interspersed with folded "shadow" bands [Fig.~\ref{fig:fig1}(e), lighter color], particularly near the X-point of the Brillouin zone. We now turn to using linearly polarized ARPES measurements to directly probe how this CDW formation reconstructs the Fermi surface.

\subsection{Reconstructed Fermi surface of LaTe$_3$ probed with linearly polarized ARPES}

{\bf [P9]} A key question is whether the CDW arises from an intra-orbital or inter-orbital nesting mechanism, where the latter couples the charge and orbital degree of freedom in the CDW order parameter and is a necessary condition for the stabilization of ferroaxial order. These two scenarios can be distinguished by examining which electronic states of the Fermi surface interact and gap out. An intra-orbital CDW gaps states of the same orbital character, whereas an inter-orbital CDW gaps states of different orbital characters~\cite{zhu2015classification,zhao2017orbital,zhang2023emergent,jiang2026direct}. We therefore performed high-resolution, polarization-dependent angle-resolved ARPES to map the orbital texture of the LaTe$_3$ Fermi surface at temperatures $T\leq82\,$K well below the CDW transition temperature. 

\begin{figure}[H]
    \centering
    \includegraphics[width=1\linewidth]{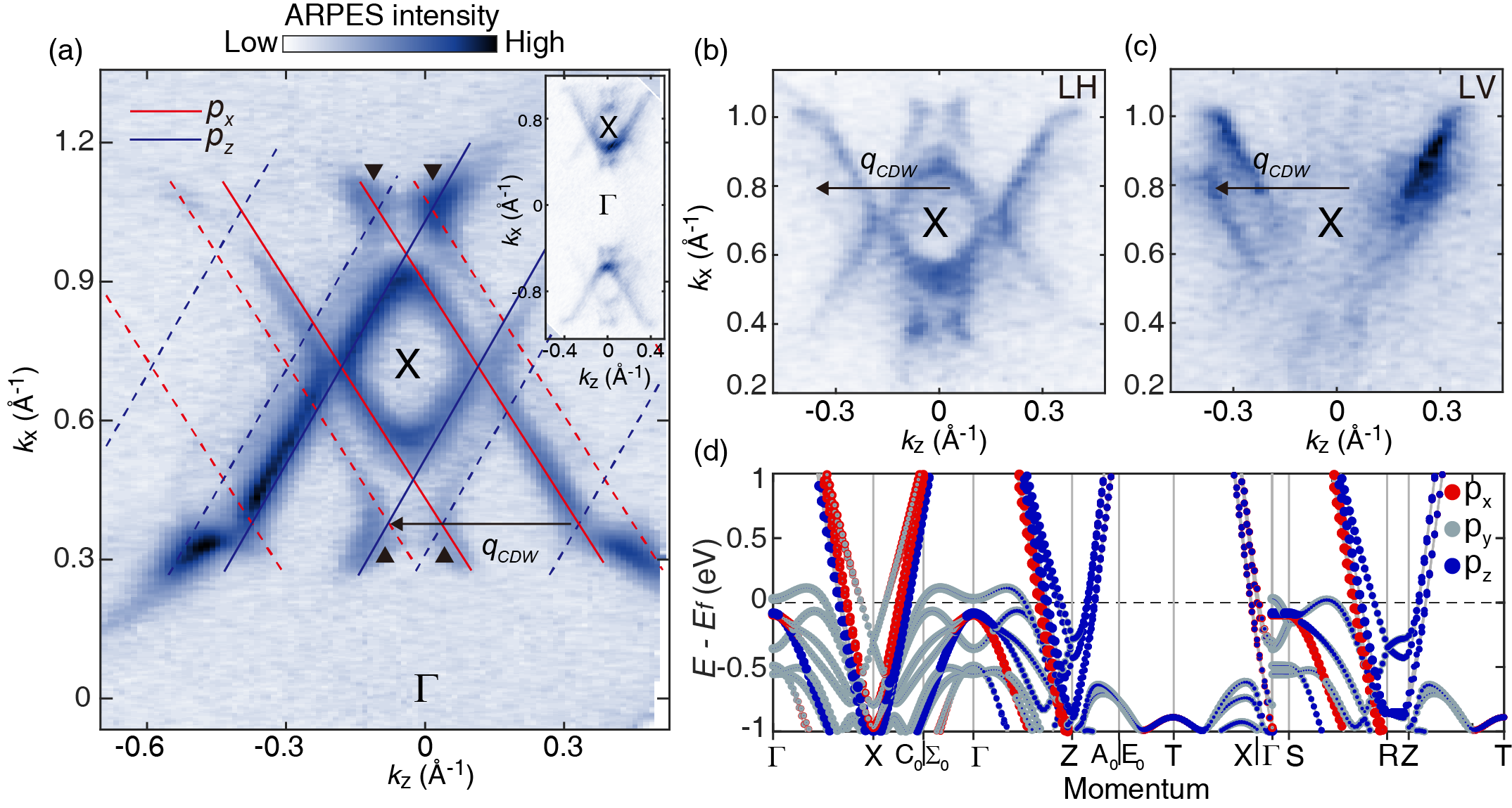}
    \caption{\textbf{Electronic structure and orbital character of LaTe$_3$ revealed by ARPES and DFT.} (a)ARPES intensity map recorded near the $X$-point of the Brillouin zone measured at a temperature $T=13\,$K with a photon energy of 38\,eV (see Methods section). The solid lines mark the original band dispersions, and the dashed lines indicate the CDW-folded replica bands. The black triangles mark the crossing points between the folded bands and the original bands, where an energy gap opens. The inset shows an ARPES map of the entire first Brillouin zone, acquired at a photon energy of $80\,$eV. (b)-(c), Shown are ARPES intensity maps acquired using linearly polarized light at $T=82\,$K with a photon energy of $77\,$eV: (b) horizontal polarization (LH) and (c) vertical polarization (LV), highlighting the orbital-selective spectral weight near the $X$-point. The unidrectional CDW wave vector $\mathbf{q}_\mathrm{CDW}$ is indicated. (d) Band structure of LaTe$_3$ in the normal state (without CDW order) obtained from density functional theory calculations (see Methods). The electronic bands are color-coded by their orbital character.}
    \label{fig:fig2}
\end{figure}

{\bf [P10]} In Fig.~\ref{fig:fig2}(a), we present an ARPES intensity map of the Fermi surface in the Brillouin zone. The Fermi surface in the vicinity of the $\Gamma$ point and along the $\Gamma Z$ path, shown in the inset, is almost entirely gapped out, which is consistent with previous studies~\cite{sarkar2023charge,brouet2008angle,smith2024uncovering} [also see Sec.~I of suppl.~materials for gap size determination]. In contrast, near the $X-$point, a complex landscape of Fermi surface features can be detected. The data reveal both the original bands with high intensity (solid lines) and their weaker, folded replicas (dashed lines) resulting from the CDW formation. At the crossing points between original and folded bands (red markers), a clear suppression of spectral weight indicates the opening of CDW gaps. Interestingly, these gaps appear at crossing points of bands with different orbital character, suggesting the presence of an inter-orbital CDW.

{\bf [P11]} To determine the orbital nature of these gapped states, we used linearly polarized ARPES to selectively highlight different orbital contributions to the Fermi surface. As shown in Figs.~\ref{fig:fig2},~(b) and (c), vertical polarization and horizontal polarization reveal distinct Fermi surface features. Through analysis of the matrix element effect, we find that for our experimental geometry adopted in this work [shown in Sec.~II of the suppl.~materials], vertical (LV) and horizontal (LH) polarization predominantly probe the $p_z$ and $p_x$ orbitals, respectively. We find that the diamond-shaped pocket at the $X-$point is mostly composed of states with $p_z$-orbital character. Crucially, the states connected by the primary CDW wave vector, $\mathbf{q}_\mathrm{CDW}$, appear to have different dominant orbital characters in these measurements, providing initial evidence for an inter-orbital CDW.

{\bf [P12]} However, in contrast to the idealized square-net Fermi surface shown in Fig.~\ref{fig:fig1}(d), where $p_x$ and $p_z$ bands are fully separated, the data in Figs.~\ref{fig:fig2},~(b) and (c) show that many bands at the Fermi surface exhibit considerable spectral weight in both polarization channels. For example, a subset of the upward dispersing bands at the $X-$point near top of the image are visible under both polarization conditions. This observation is confirmed by results from Density Functional Theory calculations of the unreconstructed Fermi surface (see Methods section) which reveal overlapping bands with $p_x$ and $p_z$ character near the $X-$point [Fig.~\ref{fig:fig2}(d)]. This complex orbital texture of the Fermi surface makes it challenging to definitively classify the CDW as purely inter- or intra-orbital from the linearly polarized ARPES data alone. This ambiguity highlights the need for a complementary real-space probe to clarify the microscopic nature of the order parameter.

\subsection{Enhancing QPI amplitude through controlled deposition of scattering centers}

\begin{figure}[H]
    \centering
    \includegraphics[width=1\linewidth]{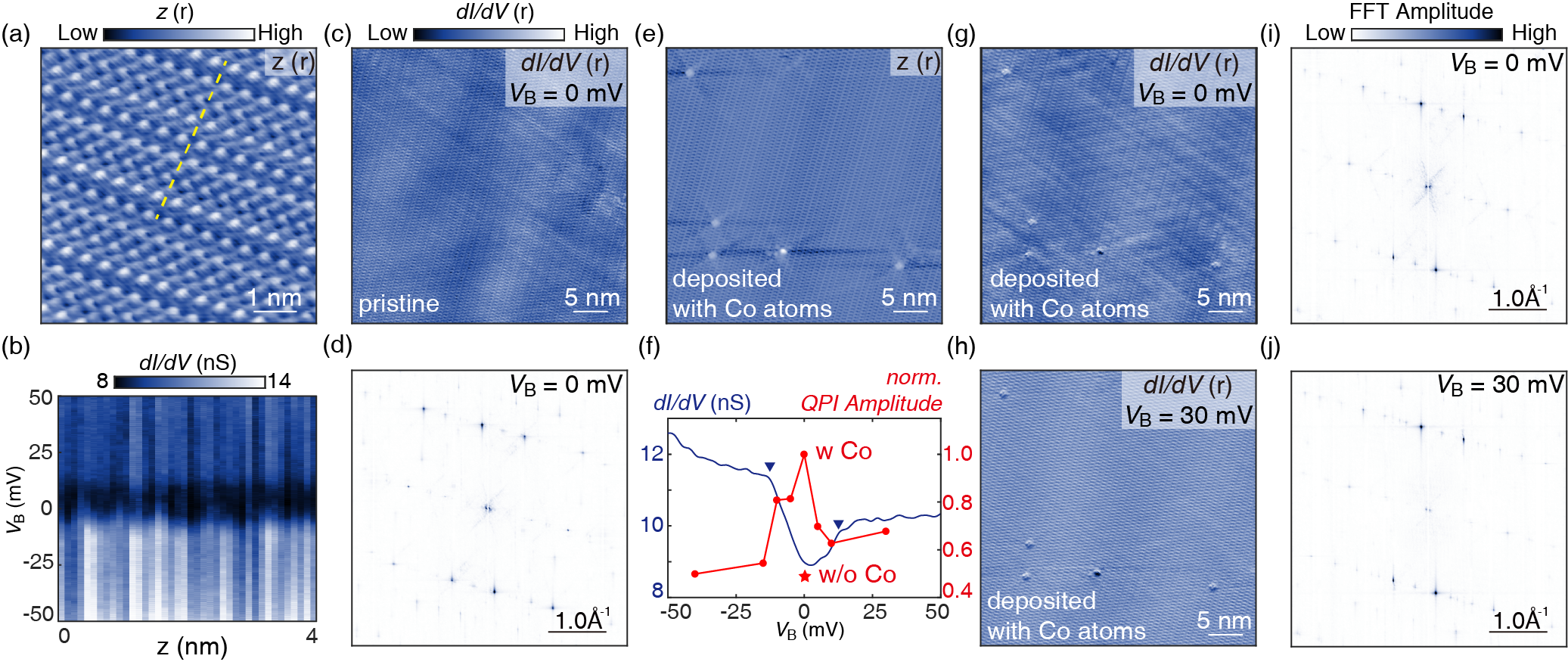}
    \caption{\textbf{Enhanced QPI pattern via depositing Co atoms.} (a) STM topography of the Te-terminated surface of LaTe$_3$ ($V_{\rm B}=300\,$mV, $I=500\,$pA, $T=4\,$K). (b) Shown is a series of $dI/dV$ spectra obtained along the yellow dashed line in panel a ($V_{\rm B}=50\,$mV, $I=1\,$nA, $V_{\rm m}=0.5\,$mV, $T=4\,$K). (c) and (d), $dI/dV$ map and corresponding two-dimensional fast-Fourier transform (2D-FFT) of the pristine Te-terminated surface measured at $V_{\rm B}=0\,$mV, respectively. (e) STM topography of the Te-terminated surface after Co atom deposition ($V_{\rm B}=80\,$mV, $I=2\,$nA, $T=4\,$K). (f) $dI/dV$ spectrum obtained on the Te-terminated surface (blue solid line, $V_{\rm B}=50\,$mV, $I=1\,$nA, $V_{\rm m}=0.5\,$mV, $T=4\,$K) and normalized QPI amplitude plotted as a function of $V_{\rm B}$. Red markers with solid line denote the QPI amplitude of $dI/dV$ maps acquired at different bias voltages on Co-deposited sample surfaces and the star represents the QPI amplitude of $dI/dV$ maps acquired at zero bias voltage on the pristine sample surface [see Sec.~IV of suppl.~materials]. (g)-(h), Shown are $dI/dV$ maps recorded on the Te-terminated surface after Co atom deposition, $V_{\rm B}=0\,$mV and $30\,$mV, respectively.(i)-(j), Shown are the 2D-FFT of the $dI/dV$ maps presented in panels g and h, respectively.}
    \label{fig:fig3}
\end{figure}

{\bf [P13]} We performed STM measurements to probe in real space how the CDW reconstructs the electronic states at the Fermi surface. Specifically, we employed QPI mapping, a technique that visualizes electronic scattering vectors between different parts of the electronic band structure, to probe the symmetry of the CDW ground state. However, the pristine cleaved surface of our high-quality LaTe$_3$ crystals presents a low density of natural defects with a seemingly low scattering cross section. This results in only weak QPI signals near Fermi energy [Figs.~\ref{fig:fig3}(c, d)] and only scattering vectors related to the structural distortion discussed in Fig.~\ref{fig:fig1}(g) are clearly visible~\cite{nakamura2024revealing}.

{\bf [P14]} To significantly enhance the intensity of the QPI signals, we evaporated cobalt (Co) atoms {\em in-situ} onto the cold ($T = 4$\,K) LaTe$_3$ surface. At this temperature, the atoms adsorb individually, acting as point-like scattering centers without forming clusters. The adsorbed Co adatoms are stable and immobile during STM measurements, allowing for the acquisition of high-resolution dI/dV maps [Fig.~\ref{fig:fig3}(e), Fig.~S2 of suppl.~materials]. Their presence dramatically sharpens the real-space interference patterns near $q=0$ compared to the pristine surface, as seen by comparing the $dI/dV$ maps at the Fermi energy ($V_{\rm B}=0\,$mV) in Fig.~\ref{fig:fig3}(g) (with Co) and Fig.~\ref{fig:fig3}(c) (pristine). The effect of the Co adatoms is most striking in the 2D-FFTs of the $dI/dV$ maps. The 2D-FFT at Fermi energy ($V_{\rm B}=0\,$mV) recorded on the Co-deposited surface [Fig.~\ref{fig:fig3}(i)] exhibits sharp, cross-like features near its center, which are barely discernible in the data recorded on the pristine surface [Fig.~\ref{fig:fig3}(d)]. Note that the 2D-FFTs are normalized to the amplitude of the $q_{\mathrm{c}}$ peak. Hence, the Co atoms may provide additional scattering channels, amplifying quasiparticle interference processes in the reconstructed Fermi surface that would be difficult to detect otherwise.

{\bf [P15]} Furthermore, by analyzing the bias voltage dependence of the QPI maps, we find that the amplitude of these cross-like QPI features is directly correlated with a region of suppressed density of states observed in tunneling spectra within $\pm20\,$mV of the Fermi energy [Fig ~\ref{fig:fig3}(b)]. The QPI patterns are significantly stronger when measured at bias voltages inside this region [Fig.~\ref{fig:fig3},~(g) and (i)] compared to outside of it [Fig.~\ref{fig:fig3},~(h) and (j)]. This bias voltage dependence of the QPI amplitude, plotted directly in Fig. ~\ref{fig:fig3}(f), suggests that these enhanced scattering vectors originate from electronic states near Fermi energy~\cite{nakamura2024revealing,xian2023coexistence} [see Sec.~IV of suppl.~materials for bias-voltage dependent $dI/dV$ maps and QPI analysis]. At these bias voltages, the CDW formation leads to a partial gapping of the Fermi surface at band crossing points, as seen in our ARPES measurements in Fig.~\ref{fig:fig2}.

\subsection{Inter-orbital CDW revealed by QPI imaging with the STM}

{\bf [P16]} We now perform a symmetry-based analysis of the QPI signal near Fermi energy and show that our results are consistent with an inter-orbital CDW phase with ferroaxial component. To this end, we plot the amplitude of the 2D-FFT recorded at Fermi energy ($V_{\rm B}=0\,$mV) from Fig.~\ref{fig:fig3}(i) on a square-root scale in Fig.~\ref{fig:fig4}(a) to further enhance the signal contrast of the cross-like QPI features near $q=0$. This 2D-FFT reveals a rich landscape of scattering vectors and branches that become even more visible in a magnified view (rotated by 60°) in Fig.~\ref{fig:fig4}(b). We will focus on two three QPI features that will be of relevance in determining the nature of the CDW. One critical feature is two side crossings of QPI branches ${\bf q}_1$ along the $c^*$ direction, centered at $q\approx\pm 0.35\,\text{\AA}^{-1}$, indicated by yellow dashed lines. the other features are two central crossings of QPI branches ${\bf q}_2$ and ${\bf q}_3$ centered near ${\bf q}=0$ indicated by red and purple dashed lines.

{\bf [P17]} To analyze these QPI characteristics, we built a minimal 2D tight-binding model on the Te square lattice within the $p_x$ and $p_z$ orbital basis of the Fermi surface. Within this $(p_x, p_z)$ orbital subspace, the CDW order parameter $\Delta_{\rm CDW}({\bf k})$ is expressed as a $2\times2$ matrix and can generally be expanded in terms of a linear combination of Pauli matrices $\sigma_i$ [$i=0,1,2,3$], $\Delta_{\rm CDW}(\mathbf{k}) = \Delta_{\rm CDW}\sum_{i=0}^3 x_i \sigma_i$ with weights $x_i$. Each Pauli matrix $\sigma_i$ implements a different coupling of electronics bands within the $(p_x, p_z)$ orbital subspace by the CDW. Therefore, the specific choices of $x_i$ encode the internal structure of the CDW by determining the symmetry breaking characteristics of its order parameter $\Delta_{\rm CDW}(\mathbf{k})$. For ferroaxial order to be present in our context, $\Delta_{\rm CDW}({\bf k})$ must break all vertical mirror symmetries $m_x,\,m_y,\,m_{xy},$ and $m_{x\bar{y}}$ [see Fig.~\ref{fig:fig1}(b)]~\cite{hu2014coexistence, alekseev2024charge,singh2025ferroaxial}. 

%{\bf [P17]} To analyze these QPI characteristics, we built a minimal 2D tight-binding model on the Te square lattice within the $p_x$ and $p_z$ orbital basis of the Fermi surface. We then calculated the QPI patterns using the $T$-matrix approach~\cite{wang2003quasiparticle} [see Methods]. In this model, the CDW order parameter $\Delta_{\rm CDW}({\bf k})$, which expressed as a $2\times2$ matrix in the $(p_x, p_z)$ orbital basis, governs the gap-opening character of band crossings at the Fermi surface. The ordering vector $\bf q_{\rm CDW}$ itself breaks the vertical mirrors $m_x$ and $m_y$ [see Fig.~\ref{fig:fig1}(b)]. For ferroaxial order to be present, the internal structure of $\Delta_{\rm CDW}({\bf k})$ thus must break the remaining $m_{xy}$ and $m_{x\bar{y}}$ symmetries. 

{\bf [P18]} We analyzed different order parameter representations in terms of their symmetry breaking characteristics [see Sec.~V of the suppl.~materials for details of symmetry analysis]. Order parameters with diagonal structure, $\Delta_{\rm CDW}({\bf k})=\Delta_{\rm CDW}\,\sigma_0$ and $\Delta_{\rm CDW}({\bf k})=\Delta_{\rm CDW}\,\sigma_z$, where $\sigma_0$ and $\sigma_z$ denote the zeroth and third Pauli matrix, respectively, define a conventional intra-orbital CDW that gaps band crossings between states of identical orbital character ($p_x$--$p_x$ or $p_z$--$p_z$). Order parameters with off-diagonal terms of type $\Delta_{\rm CDW}({\bf k})=\Delta_{\rm CDW}\,\sigma_{i}$, where $\sigma_{i}$ denote the Pauli matrices ($i=x,\,y$), define an inter-orbital CDW that gaps band crossings of $p_x$- and $p_z$-derived bands with order parameter amplitude $\Delta_{\rm CDW}$. Critically, our symmetry analysis shows that not all inter-orbital CDWs have a ferroaxial order parameter. Only inter-orbital CDWs of structure $\Delta_{\rm CDW}({\bf k})=\Delta_{\rm CDW}\,i\sigma_{y}$ break all mirror symmetries and thus implements ferroaxial order, whereas $\Delta_{\rm CDW}({\bf k})=\Delta_{\rm CDW}\,\sigma_{x}$ only affects $m_{x}$ and $m_{y}$ which are already broken by the ordering vector $q_{\rm CDW}$ of the CDW~\cite{singh2025ferroaxial}. 
As $\sigma_x$ and $i \sigma_y$ belong to different symmetry representations, they do not develop simultaneously at the transition point $T=T_{\rm CDW}$. However, as the ferroaxial $i \sigma_y$ order breaks all vertical mirror symmetries, deep inside the CDW phase it generally seeds a mixed phase  $\Delta_{\rm CDW}({\bf k})=\Delta_{\rm CDW}\,[(1-x)\sigma_x+xi\sigma_{y}]$ at $T<T_{\rm CDW}$, where the parameter $x$ determines the relative weight of the $i\sigma_y$ channel. In contrast, if the primary order developed at $T_{\rm CDW}$ was $\sigma_x$, then a mixed order would not emerge without additional symmetry breaking transition.
% Furthermore, mixed order parameter structures can generally exist as weighted linear combinations of $\sigma_i$ below the ordering temperature $T_{\rm CDW}$. For example, because $i\sigma_y$ breaks all vertical mirror symmetries, a ferroaxial CDW can generally seed a mixed phase Note that symmetry, however, prevents a direct transition into such mixed phase at $T=T_{\rm CDW}$, because $\sigma_x$ and $i\sigma_y$ belong to different symmetry classes.

{\bf [P19]} Moreover, the Co adatoms independently introduce local scattering potentials $\hat{V}$ whose orbital structure determines which quasiparticle channels they couple: a diagonal potential ($\hat{V}\propto\sigma_0$, intra-orbital impurity scattering) confines scattering within a single orbital channel, whereas an off-diagonal potential ($\hat{V}\propto\sigma_x$, inter-orbital impurity scattering) admixes $p_x$ and $p_z$ quasiparticle states. Because the coherence factor $\langle u_m(\mathbf{k})|\hat{V}|u_n(\mathbf{k}+\mathbf{q})\rangle$ in the $T$-matrix expression (see Methods) is appreciable only when the orbital symmetry of $\hat{V}$ is compatible with the orbital mismatch between the initial state $|u_m\rangle$ and the final state $|u_n\rangle$, each combination of CDW type and impurity type selectively enhances a distinct subset of scattering wavevectors, imprinting an orbital-resolved fingerprint directly onto the QPI characteristics. 

{\bf [P20]} We have simulated QPI patterns using the different CDW order parameter structures, impurity potentials, and order parameter amplitudes using the $T$-matrix approach~\cite{wang2003quasiparticle} [see Methods]. Our detailed analysis of the simulated QPI patterns [see Sec.~V of suppl.~materials] reveals that simulated QPI patterns of a mixed phase $\Delta_{\rm CDW}({\bf k})=\Delta_{\rm CDW}\,[0.6\sigma_x+0.4i\sigma_{y}]$ with $\Delta_{\rm CDW}=0.2$ [Fig.~\ref{fig:fig4},~(c) and (d)] most accurately match the experimentally observed QPI patterns shown in Fig.~\ref{fig:fig4}(b) among all possible order parameters that can exist within the $p_x-p_z$ orbital subspace of the Fermi surface. When only intra-orbital quasiparticle scattering ($p_x$--$p_x$ or $p_z$--$p_z$) is permitted within this phase, the two side-crossings ${\bf q}_1$ (yellow dashed lines) seen in our experimental data are captured. When inter-orbital scattering processes ($p_x$--$p_z$ or $p_z$--$p_x$) are also included, the simulated QPI pattern reproduces the two central QPI crossings ${\bf q}_2$ and ${\bf q}_3$ (red and purple dashed lines). 

{\bf [P21]} On the other hand, in the limit $x\rightarrow0$, a conventional inter-orbital CDW ($\sigma_x$) is realized, which does not break all vertical mirror symmetries. Its simulated QPI pattern [Fig.~\ref{fig:fig4},~(e) and (f)] features a halo-like structure near $q=0$ from inter-orbital scattering that connects ${\bf q}_1$ with ${\bf q}_{2,3}$. This feature is not present in our experimental data [Fig.~\ref{fig:fig4}(b)]. Moreover, simulated QPI patterns based solely on quasiparticle scattering in an intra-orbital CDW state, $\Delta_{\rm CDW}({\bf k})=\Delta_{\rm CDW}\,\sigma_0$ and $\Delta_{\rm CDW}({\bf k})=\Delta_{\rm CDW}\,\sigma_z$, are dominated by only a single crossing or a complex pattern of QPI branches depending on the scattering channel [Fig.~\ref{fig:fig4},~(g-h) and Fig.~S12], also failing to reproduce these experimentally observed QPI features.

{\bf [P22]} Furthermore, when comparing the experimentally detected QPI characteristics [Fig.~\ref{fig:fig4}(b)] with the Fermi surface measured using ARPES, we can directly identify the different quasiparticle scattering vectors based on our experimental data alone. To this end, we represent the QPI branches ${\bf q}_{1,2,3}$ as vectors whose lengths and directions are determined by the experimentally detected QPI pattern. We then overlay these vectors on the measured Fermi surface near the $X-$point to identify the scattering processes between different parts of the Fermi surface that match these vectors [Fig.~\ref{fig:fig5},~(a)-(c)]. We find that ${\bf q}_{1}$, ${\bf q}_{2}$, and ${\bf q}_{3}$ connect electronic states near gapped-out band crossing points, which are formed by $p_x$ and $p_z$ bands, with other parts of the Fermi surface within the reconstructed Brillouin zone. This finding is consistent with these vectors arising from scattering within an inter-orbital CDW phase in our QPI simulations [Fig.~\ref{fig:fig4},~(c) and (d)], where the inter-orbital CDW causes a characteristic reconstruction of the Fermi surface. It is interesting to note that ${\bf q}_{3}$, which describes inter-orbital scattering in an inter-orbital CDW phase, arises from scattering between overlapping $p_x$ and $p_z$ bands, previously detected in our linearly polarized ARPES measurements and DFT calculations [Fig.~\ref{fig:fig2},~(b)-(d)].

\subsection{Discussion and Conclusion}

\begin{figure}[H]
    \centering
    \includegraphics[width=1\linewidth]{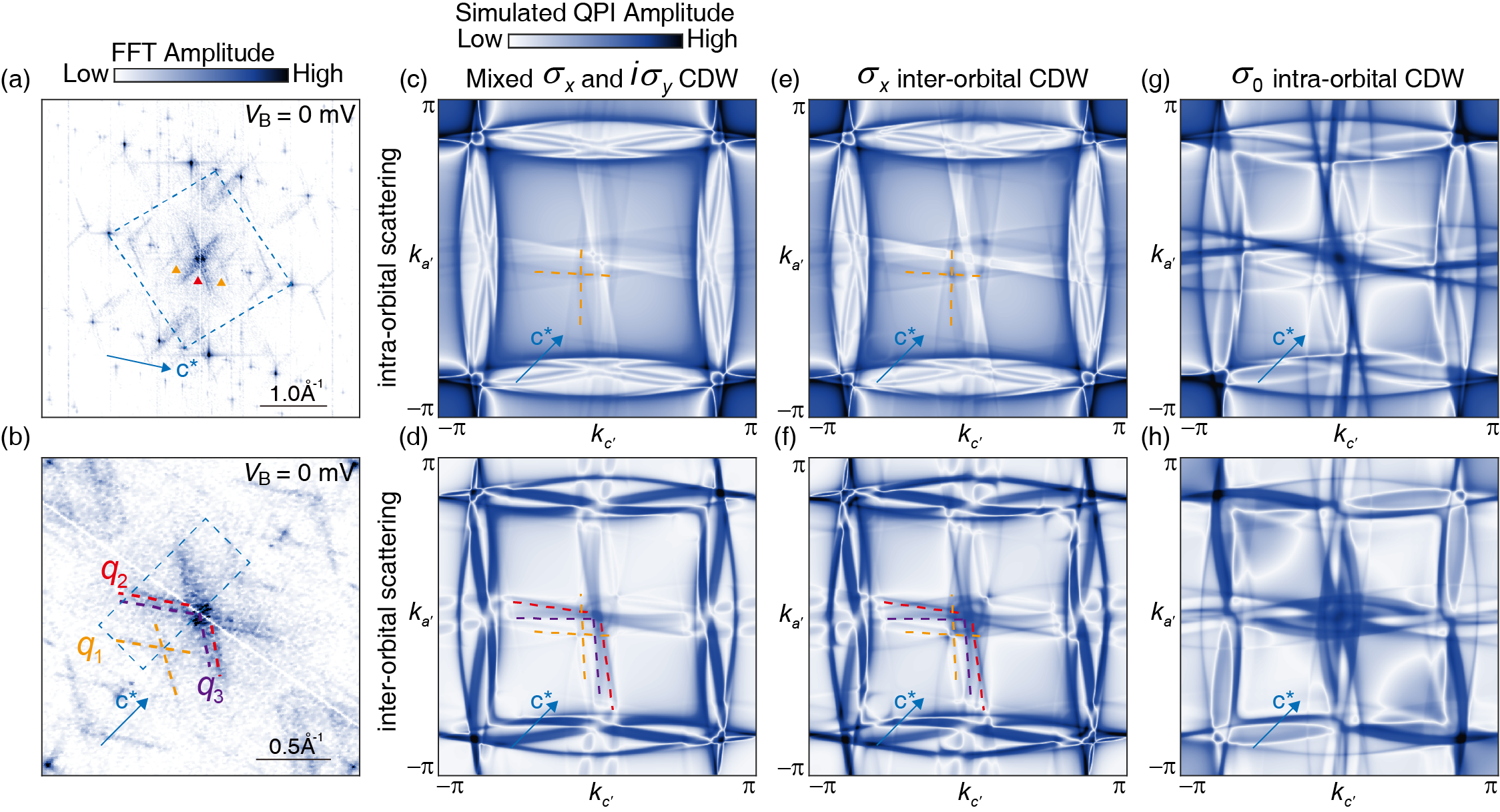}
    \caption{\textbf{Evidence of an inter-orbital CDW in LaTe$_3$ from quasiparticle interference.} (a) Shown is the two-dimensional fast Fourier transform (2D-FFT) of the $dI/dV$ map at Fermi energy ($V_{\rm B}=0\,$mV) in Fig.~\ref{fig:fig3}(g), here, plotted on a square-root color scale. (b) Zoomed-in view of the blue dashed box in panel a, showing the QPI pattern within the two-Te Brillouin zone. The different quasiparticle scattering branches $q_1$, $q_2$, and $q_3$ are indicated by color-coded markers (in panel a) and dashed lines. (c)-(d) Shown are tight-binding simulated QPI patterns for  an inter-orbital CDW with ferroaxial order (0.6$\times$$\sigma_x$+0.4$\times$$i\sigma_y$), (e)-(f) an inter-orbital CDW ($\sigma_x$), and (g)-(h) a conventional intra-orbital CDW ($\sigma_0$) for both intra- (top row) and inter-orbital (bottom row) impurity scattering channels.}

    \label{fig:fig4}
\end{figure}

\begin{figure}[H]
    \centering
    \includegraphics[width=1\linewidth]{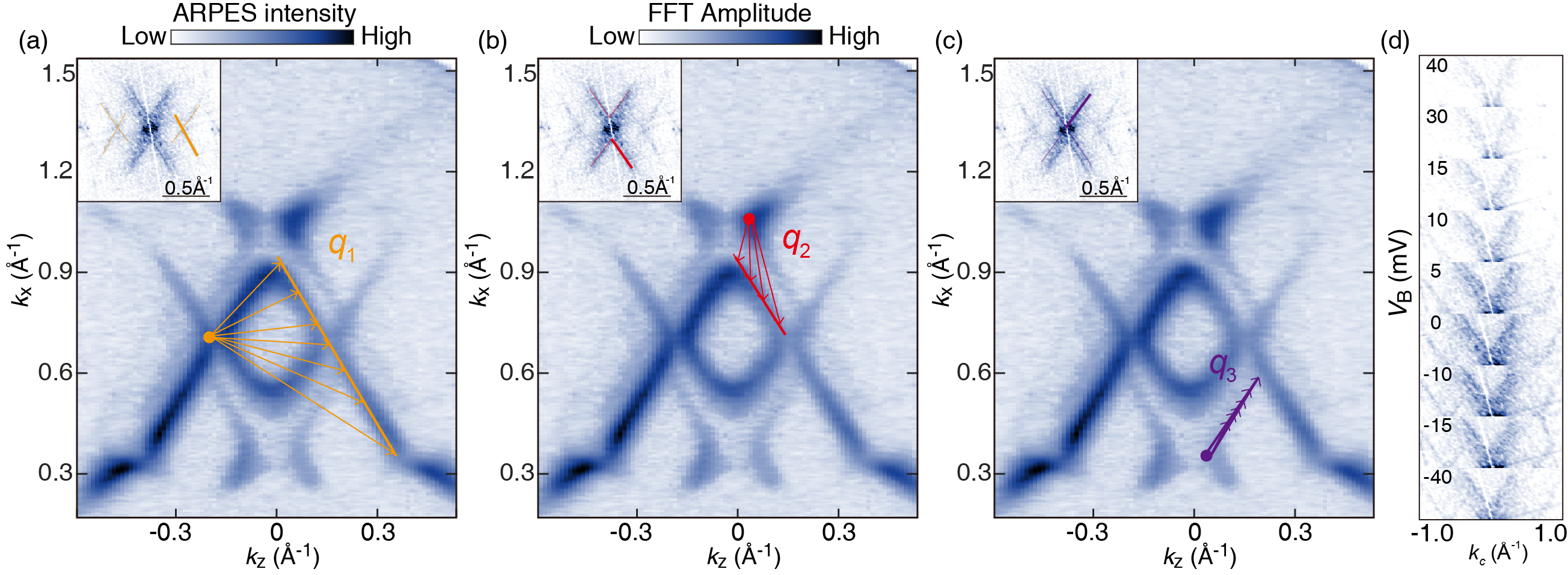}
    \caption{\textbf{Quasiparticle scattering vectors at the Fermi surface detected in STM and ARPES measurements.} (a-c) Shown are the scattering vectors ${\bf q}_1$, ${\bf q}_2$, and ${\bf q}_3$ extracted from the QPI maps measured with STM overlaid on the Fermi surface near the $X-$point of the Birllouin zone measured with ARPES ($T=13\,$K with a photon energy of $38\,$eV). Yellow scattering vectors correspond to the two side crossings (${\bf q}_1$), while red (${\bf q}_2$) and purple (${\bf q}_3$) vectors represent two distinct sets of QPI branches associated with the central crossing. The vectors and Fermi surface portions marked in the ARPES data only correspond to the bold features in the QPI pattern. (d) Shown is a series of magnified QPI patterns as a function of bias voltage. The reciprocal space region corresponds to the region marked by the dashed blue boxes in Fig.~\ref{fig:fig4}(b). The corresponding $dI/dV$ maps are shown in Sec.~IV of the suppl.~materials}
    \label{fig:fig5}
\end{figure}

{\bf [P23]} Our analysis of the measured QPI maps [Fig.~\ref{fig:fig4}] combined with the identification of the QPI vectors in the Fermi surface measured with ARPES [Fig.~\ref{fig:fig5}] suggests that the order parameter of the unidirectional CDW in LaTe$_3$ has a ferroaxial component [$i\sigma_y$]~\cite{singh2025ferroaxial}. A pure conventional inter-orbital CDW [$\sigma_x$], recently reported for ErTe$_3$~\cite{freitas2026revealing} cannot fully account for the QPI patterns detected in our measurements. Moreover, the presence of an intra-orbital CDW order parameter [$\sigma_0$ and $\sigma_z$] discussed to be present in CeTe$_3$~\cite{smith2024uncovering} can also be ruled out according to our analyses. In this regard, bias voltage-dependent QPI characteristics near the $X-$point, shown in Fig.~\ref{fig:fig5}(d), also reveal a pronounced non-equivalence of the scattering patterns above and below the Fermi level [$dI/dV$ maps are shown in Sec.~IV of the suppl.~materials]. This asymmetry is consistent with the particle--hole-asymmetric multiband electronic structure of LaTe$_3$, which highlights the role of orbital degrees of freedom for the CDW formation~\cite{zhao2017orbital} in a multi-band nesting picture~\cite{nowadnick2012quasiparticle,gruner2018density}. The noticeable weakening of QPI intensity detected at $|V_{\rm B}|>20\,$mV [also see Fig.~\ref{fig:fig3}(f)] is consistent with our scattering vector analysis of Fig.~\ref{fig:fig5} according to which QPI involves the diamond-shaped electron pocket centered at the $X-$point that exists over a narrow energy range near Fermi energy~\cite{nakamura2024revealing}.

{\bf [P24]} In summary, by combining orbital-resolved momentum- and real-space spectroscopic techniques, we provide unique insight into the electronic fingerprint of the CDW in LaTe$_3$. Our study provides experimental evidence for an inter-orbital order parameter that couples charge and orbital degrees of freedom~\cite{hu2014coexistence, alekseev2024charge,singh2025ferroaxial}. A symmetry-guided analysis of QPI maps measured at Fermi energy suggests the presence of a ferroaxial order parameter component, consistent with reports based on optical measurements of collective mode excitations~\cite{singh2025ferroaxial}. Our results demonstrate that direct spectroscopic characterization of the electronic structure of CDW order parameters, as performed in our study, is critical to ultimately determine their underlying symmetry and character. In this context, a detailed analysis of the scattering potential induced by the adsorbed Co atoms suggests that their presence can open additional inter-orbital quasiparticle scattering channels [Sec.~VI of the suppl.~materials], helping to resolve the orbital structure of the CDW order parameter. Hence, our work establishes a comprehensive spectroscopic pathway for precisely identifying complex charge order parameters using QPI with STM.

{\bf [P25]} The insights gained here open critical future avenues for the study of quantum materials hosting intertwined orders. It will be of immediate interest to shed more light on the evolution of a pure ferroaxial phase $\Delta_{\rm CDW}({\bf k})=\Delta_{\rm CDW}\,i\sigma_y$ reported to form at $T=T_{\rm CDW}$~\cite{singh2025ferroaxial} into the mixed phase $\Delta_{\rm CDW}({\bf k})=\Delta_{\rm CDW}\,[(1-x)\sigma_x+xi\sigma_{y}]$ with best fit to data for $x=0.4$] detected in our study at 4\,K using temperature-dependent measurements of heavier tritellurides, such as ErTe$_3$ or HoTe$_3$. Temperature and strain-dependent measurements~\cite{freitas2026revealing,guo2024correlated} could also shed light on the role of domain and strain effects on the presence and electronic signatures of the unconventional CDW order in these materials. It is also worth mentioning that pressure-induced superconductivity has recently been reported to compete with the CDW state in LaTe$_3$~\cite{wang2026pressure}. The identification of a ferroaxial CDW component in our study could be highly relevant to understanding the microscopic pairing mechanism underlying this newly observed superconducting state. Ultimately, our findings refine the search for hidden orders governed by charge-lattice-orbital interactions, paving the way for their eventual optical control~\cite{zeng2025photo} for optoelectronic applications above-room-temperature.

\clearpage
\bibliography{bibliography}

\clearpage
\section{Figures}
%\clearpage

\clearpage
\section{Methods}

\subsubsection{Synthesis of LaTe$_3$ crystals}

Single crystals of LaTe$_3$ were grown in excess tellurium using the self-flux method where Te (metal basis 99.999, Sigma-Aldrich) was mixed with lanthanum (99.9$\%$, Sigma-Aldrich) in a mass ratio of 97:3. The reagents were sealed in a quartz ampule under vacuum and repeatedly evacuated and backfilled with Argon gas. The samples were heated to 900°C over a period of 12h, held at that temperature for two days, then cooled down to 550°C at a rate of 2°C/h. The crystals were separated from the flux via centrifugation at 550°C. The excess flux was then removed by chemical vapor transport. The centrifuged quartz ampule was placed in a tube furnace with the crystals facing the heat source and heated to 420°C over six hours, held for five days, and then cooled to room temperature over six hours.

\subsubsection{Scanning Tunneling Microscopy (STM) Measurements}

The LaTe$_3$ samples were cleaved after cooling down to a temperature $T=4.3\,$K inside an ultra-high vacuum (UHV) chamber with a base pressure of $p\approx1.4 \times 10^{-10}\,$mbar. Several crystals of doped and undoped LaTe$_3$ were cleaved and the results presented in this manuscript were consistently observed. STM measurements were conducted using a home-built STM instrument under cryogenic ($T=4.3\,$K) and UHV ($p\approx1.4 \times 10^{-10}\,$mbar) conditions using a chemically etched tungsten tip. The tip was prepared on a Cu(111) surface through field emission and controlled indentation, as well as calibrated against the Cu(111) Shockley surface state before each set of measurements. Bias voltage ($V$) dependent differential conductance ($dI/dV$) spectra and maps were recorded using standard lock-in methods with a bias modulation $1\,\text{mV}\leq V_{\rm m}\leq10\,\text{mV}$ at a frequency $f=719.7\,$Hz, as indicated in the main text. The $dI/dV$ maps were recorded using multi-pass mode to avoid set-point effects. To enhance the clarity of the data presentation, high-frequency noise originating from mechanical vibrations coupling to the tip-sample junction was removed from the raw data and interpolation was performed to reduce pixelated appearance where applicable.

\subsubsection{Angle-resolved photo-emission spectroscopy measurements}

The LaTe$_3$ single crystals were cleaved in situ under an ultra-high vacuum (UHV) condition with a base pressure better than $5 \times 10^{-11}$ torr. Regular ARPES measurements were carried out at the BL03U end station of the Shanghai Synchrotron Radiation Facility (SSRF), using a Scienta-Omicron DA30L electron analyzer. The overall energy resolution was set better than 20 meV, and the angular resolution was better than 0.1$^\circ$ during the measurements. Polarization-dependent ARPES experiments were performed at the TPS 39A1 end station of the National Synchrotron Radiation Research Center (NSRRC), which was also equipped with a Scienta-Omicron DA30L analyzer.

\subsubsection{Density functional theory calculation}
Density Functional Theory calculations of LaTe$_3$ were performed using the Vienna {\em ab-initio} simulation package (VASP)~\cite{kresse1993ab}. The exchange-correlation functional was described by the generalized gradient approximation (GGA) with the Perdew-BurkeErnzerhof (PBE) functional type~\cite{blochl1994projector,perdew1996generalized}. The cutoff energy for plane-wave basis was set to 600 eV and the BZ was sampled by $9\times3\times9$ $\Gamma$-centered k mesh. The convergence criterion for the energy in the self-consistent-field cycle and total force tolerance on each atom were set to $10^{-6}$ eV and 0.02 eV/\AA{}, respectively.

\subsubsection{Tight-binding model calculations}

We implement a tight-binding Hamiltonian $H_0(\mathbf{k},\mu)$ of the square-net electronic structure as a $2\times2$ matrix to account for the orbital degree of freedom:
\[
H_0(\mathbf{k}) = 
\begin{pmatrix}
t_\sigma \cos k_x - t_\pi \cos k_y - \mu & 2t_d \sin k_x \sin k_y \\
2t_d \sin k_x \sin k_y & t_\sigma \cos k_y - t_\pi \cos k_x - \mu
\end{pmatrix}
\]
where $t_\sigma$, $t_\pi$, and $t_d$ are orbital-dependent hopping amplitudes,  $\mu$ is the chemical potential.

The unidirectional charge density wave (CDW) with order parameter $\Delta_{\rm CDW}$ is incorporated by extending the $H_0(\mathbf{k})$ ($2\times2$ matrix) to a Hamiltonian involved with CDW~\cite{wang2003quasiparticle}.
\[
H_{\rm CDW}(\mathbf{k}) = 
\begin{pmatrix}
H_0(\mathbf{k},\mu) & \Delta_{\rm CDW} & \Delta_{\rm CDW}^\dagger \\
\Delta_{\rm CDW}^\dagger & H_0(\mathbf{k}+\mathbf{Q}_{\rm CDW},\mu) & \Delta_{\rm CDW} \\
\Delta_{\rm CDW} & \Delta_{\rm CDW}^\dagger & H_0(\mathbf{k}+2\mathbf{Q}_{\rm CDW},\mu)
\end{pmatrix}.
\]
We approximate the incommensurate CDW of LaTe$_3$ by choosing the nearby commensurate ordering wave vector $\mathbf{Q}_{\rm CDW}=(4\pi/3,4\pi/3)$ [see Sec.~VII of suppl.~materials for details].

We implement intra- and inter-orbital CDW/impurity scattering as follows: Intra-orbital CDW scattering is captured by the diagonal blocks of the $\Delta_{\rm CDW}(\mathbf{k})$, while inter-orbital CDW scattering is mediated by the off-diagonal CDW order parameter $\Delta_{\rm CDW}$. For impurities, intra-orbital scattering uses a diagonal potential $\hat{V}_{\rm intra}$, and inter-orbital scattering uses an off-diagonal potential $\hat{V}_{\rm inter}=\sigma_x$.

 Now we describe in detail how each type of scattering is encoded in the model. The CDW order parameter $\Delta_{\rm CDW}$ is a $2\times2$ matrix in the $(p_x,p_z)$ orbital space. An intra-orbital CDW is represented by a diagonal matrix
\[
\Delta_{\rm CDW}^{\rm intra} = \begin{pmatrix}\Delta_{\rm CDW} & 0 \\ 0 & \Delta^\prime_{\rm CDW}\end{pmatrix},
\]
which independently gaps the $p_x$--$p_x$ and $p_z$--$p_z$ band crossings, leaving the $p_x$--$p_z$ hybridization gap absent. An inter-orbital CDW is instead represented by an off-diagonal matrix such as
\[
\Delta_{\rm CDW}^{\rm inter} = \begin{pmatrix}0 & \Delta_{\rm CDW} \\ \Delta_{\rm CDW}^\prime & 0\end{pmatrix},
\]
which couples $p_x$ and $p_z$ states and opens a gap at crossings between bands of different orbital character. The finite off-diagonal element corresponds physically to a CDW condensate that carries a net orbital angular momentum, breaking the equivalence of the two Te $p$-orbital channels---the defining feature of a ferroaxial order. Each of the four panels in Fig.~\ref{fig:fig4}(c)--(f) uses a single, pure order parameter (either $\Delta_{\rm CDW}^{\rm intra}$ or $\Delta_{\rm CDW}^{\rm inter}$) together with a single impurity type, allowing the contribution of each channel to be identified unambiguously.

For the impurity potential, the Co adatom is modeled as a local on-site potential $\hat{V} = v_0\,M$ placed at a single Te site, where $M$ is a $2\times2$ matrix in orbital space and $v_0$ is the scattering strength. An intra-orbital impurity has $M = \sigma_{0,z}$, i.e.\ a potential that is diagonal in orbital space: it scatters a $p_x$ electron into another $p_x$ state and a $p_z$ electron into another $p_z$ state, without mixing the two channels. An inter-orbital impurity has $M = \sigma_{x,y}$, an off-diagonal potential that flips the orbital index upon scattering ($p_x\leftrightarrow p_z$). The QPI signal is then computed via the $T$-matrix formalism as~\cite{wang2003quasiparticle}
\[
\delta\rho(\mathbf{q},\omega)=
-\frac{1}{\pi}\,\mathrm{Im}\,\mathrm{Tr}\sum_{\mathbf{k}}
G_0(\mathbf{k},\omega)\,\hat{V}\,G_0(\mathbf{k}+\mathbf{q},\omega),
\]
where $G_0(\mathbf{k},\omega)=[\omega-H_{\rm CDW}(\mathbf{k})+i\eta]^{-1}$ is the unperturbed retarded Green's function. The coherence factor $\langle u_m(\mathbf{k})|\hat{V}|u_n(\mathbf{k}+\mathbf{q})\rangle$ selects which scattering channels contribute: because the orbital content of the CDW-reconstructed eigenstates $|u_m\rangle$ varies strongly across the Fermi surface, only impurity potentials whose orbital symmetry matches the orbital mismatch between initial and final states produce a strong QPI peak at the corresponding $\mathbf{q}$ vector. 

Concretely, the lateral crossings (yellow dashed lines in Fig.~\ref{fig:fig4}(b)) are enhanced when the impurity potential is intra-orbital ($M=\sigma_0$), probing states connected by $\mathbf{q}_{\rm CDW}$ that share a dominant orbital character. The central crossings (red and purple dashed lines) are instead enhanced by an inter-orbital impurity ($M=\sigma_x$), because the states they connect carry predominantly opposite orbital character in the inter-orbital CDW background. The distinct patterns produced by these two impurity types therefore provide a direct, symmetry-resolved fingerprint of the orbital content of the CDW order parameter.

\clearpage
\section{Acknowledgments}

The authors appreciate valuable discussions with Kenneth Burch. This work was primarily supported by the Hong Kong Research Grants Council (Grant Nos.\,26304221, 16302422, 16302624, and C6033-22G awarded to BJ) and the Croucher Foundation (Grant No.\,CIA22SC02 awarded to BJ). JM acknowledges funding by the Hong Kong Research Grants Council (21304023) and the National Natural Science Foundation of China (12422405). BJ and JM acknowledge funding by the NSFC/RGC Collaborative Research Scheme (CRS\_CityU101/25). BJ and HCP acknowledge support by the Hong Kong Research Grants Council through Grant No.\ AoE/P-604/25-R. The work of ZZ and HCP are supported by the Future Science Awards Foundation (FSAFL25SC01) and the Croucher Foundation (CIA23SC01). L.M.S. acknowledges support from the AFOSR (FA9550-24-1-0110, via Boston College) and the Gordon and Betty Moore Foundation's EPiQS Initiative (GBMF9064).

\section{Author Contributions}

JZ and BJ initiated the project. JZ carried out the STM measurements with the help of LL and analyzed the data. ZZ performed the theoretical model calculations with the help of JZ. FY, ZG, TG, SH, XL, and ZL performed the angle-resolved photoemission spectroscopy measurements and performed the DFT calculations. JL synthesized the single crystals. BJ, HCP, JM, and LMS supervised the study. All authors discussed the result and contributed to the manuscript, which was written by JZ and BJ.

\section{Competing Interest Declaration} The authors declare that they have no competing financial interest.

\section{Data Availability Statement} Replication data for this study can be accessed on Zenodo via the link XXX.

\end{document}